\documentclass[twocolumn,nofootinbib,prd,aps,superscriptaddress,tightenlines]{revtex4}
\usepackage{graphicx}
\usepackage{epsfig}
\usepackage{amsmath}

\newif\ifpdf
\ifx\pdfoutput\undefined
\pdffalse % we are not running PDFLaTeX
\else
\pdfoutput=1 % we are running PDFLaTeX
\pdftrue
\fi

\def\OMIT#1{{}}
\def\lqcd{\Lambda_{\rm QCD}}

\def\mDbar{\overline{m}_D}

\def\Ecut{{E_{\rm cut}}}

\def\l#1{\ensuremath{\lambda_{#1}}}
\def\r#1{\ensuremath{\rho_{#1}}}
\def\t#1{\ensuremath{{\cal T}_{#1}}}

\def\mqhat{\ensuremath{\hat m_c}}

\newcommand{\nn}{\nonumber}
\newcommand{\beq}{\begin{equation}}
\newcommand{\eeq}{\end{equation}}
\newcommand{\beqa}{\begin{eqnarray}}
\newcommand{\eeqa}{\end{eqnarray}}

\begin{document}
%%%%%%%%%%%%%%%%%%%%%%%%%%%%%%%%%%%%%%%%%%
%Some more stuff to get graphics to work
\ifpdf
\DeclareGraphicsExtensions{.pdf, .jpg}
\else
\DeclareGraphicsExtensions{.eps, .jpg}
\fi
%%%%%%%%%%%%%%%%%%%%%%%%%%%%%%%%%%%%%%%%%%

\preprint{ \vbox{\hbox{UCSD/PTH 02-28}
  \hbox{hep-ph/0212164} }}

\title{Detector resolution effects on hadronic mass moments 
in  $B\rightarrow X\ell\nu$}

\author{Christian W.~Bauer}
\email{bauer@physics.ucsd.edu}
\affiliation{Department of Physics, University of California at San Diego,
  La Jolla, CA 92093\vspace{4pt} }

\author{Benjamin Grinstein}
\email{bgrinstein@ucsd.edu}
\affiliation{Department of Physics, University of California at San Diego,
  La Jolla, CA 92093\vspace{4pt} }

\begin{abstract}

We show how to relate moments of the reconstructed hadronic invariant
mass distribution to the moments of the physical spectrum in a model
independent way. The only information needed on detector resolution
functions are their first few moments. Theoretical predictions for the
first three moments as functions of non-perturbative parameters are
given for various lepton energy cuts.

\end{abstract}

\maketitle

Differential distribution for inclusive decay rates of heavy mesons
can be calculated using the operator product expansion (OPE)
\cite{ope}. The resulting decay distributions are
singular and can only be compared with experimentally measured
distributions after smearing them over a sufficiently large
interval. For example, calculating the hadronic invariant mass
distribution for the semileptonic decay $B \to X_c \ell \bar\nu$ using
the OPE, one finds to leading order in $\Lambda_{\rm QCD}/m_b$
\cite{fls}
\begin{eqnarray}\label{simplediff}
\frac{d\Gamma}{ds} = 
\Gamma \, \delta(s- \mDbar^2) 
        + {\cal O}\left(\frac{\Lambda_{\rm QCD}}{m_b}\right)\,,
\end{eqnarray}
where $s = m_X^2$ is the square of the hadronic invariant mass,
$\mDbar=\frac14(m_D+3m_{D^*})$ is the spin averaged $D$-$D^*$ mass,
and $\Gamma$ denotes the total semileptonic width. Non-perturbative
corrections to this differential rate yield even more singular
terms. To compare this distribution with the smooth spectrum measured
experimentally, one has to find observables which sufficiently smear
the differential spectrum. The simplest such observables are moments
of this spectrum, defined as
\begin{eqnarray}\label{Mn}
{\cal M}_n = \frac{\int \! ds \, s^{n} \frac{d\Gamma}{ds}}{\int \! ds
\, \frac{d\Gamma}{ds}}\,.
\end{eqnarray}
{}From Eq.~(\ref{simplediff}) on obtains trivially
\begin{eqnarray}
{\cal M}_n = \mDbar^{2n}\left[ 1 
  + {\cal O}\left(\frac{\Lambda_{\rm QCD}}{m_b}\right)\right]\,.
\end{eqnarray}
The corrections to the first two moments have been calculated to order
$1/m_b^3$ and $\alpha_s^2 \beta_0$ \cite{fls,fl,bllm}. By measuring
deviations from the above relation, one can therefore obtain
information about the size of the non-perturbative matrix elements. In
turn, information on these non-perturbative parameters can be used in
precise determination of CKM elements $|V_{cb}|$ and  $|V_{ub}|$.

The first two moments were first measured by the CLEO collaboration
with a lower cut on the lepton energy, to suppress the background in
which the lepton originates from a semileptonic decay of the $D$ meson
\cite{cleosh}.  More recently, the DELPHI collaboration presented
measurements of these moments without a cut on the lepton energy
\cite{delphish}, while the BaBar collaboration measured the first
moment as a function of the lepton energy cut \cite{babarsh}. This
dependence on the lepton energy cut can be compared with predictions
from theory, allowing for a more detailed test of the operator product
expansion.

In general, the reconstructed hadronic invariant mass spectrum $d
\Gamma/ds_R$ is a convolution of the true spectrum $d \Gamma/ds_T$ and
a detector resolution function $P(s_R-s_T,s_T)$, which describes the
probability of measuring the reconstructed invariant mass $s_R$,
given the true mass $s_T$
\begin{eqnarray}\label{genconv}
\frac{d\Gamma}{d s_R}\Big | _\Ecut &=& \int \! ds_T \left[
P_D(s_R-s_T,s_T,\Ecut) \frac{d\Gamma_{\!D}}{d s_T} \Big | _\Ecut
\right.\nn\\ && \left.\,\,\, + P_{D^*}(s_R-s_T,s_T,\Ecut)
\frac{d\Gamma_{\!D^*}}{d s_T}\Big | _\Ecut \right.\nn\\ &&
\left.\,\,\, + P_X(s_R-s_T,s_T,\Ecut) \frac{d\Gamma_{\!X}}{d s_T}\Big
| _\Ecut \right]\,,\,\,\,
\end{eqnarray}
where $X$ denotes any charmed final state not equal to a single $D$ or
$D^*$. In (\ref{genconv}) we have allowed for different resolution
functions for $D$, $D^*$ and $X$ final states as is expected
experimentally. Each term in this expression depends explicitly on the
value of the lepton energy cut $\Ecut$. In the remainder of this
paper, this dependence will be suppressed, however all equations should
be understood as having this dependence.

{}From the result of \cite{fls,fl,bllm} for the first two moments of
the differential decay spectrum
\begin{eqnarray}
\frac{d \Gamma}{ds} = \frac{d \Gamma_{\!D}}{ds} + \frac{d
\Gamma_{\!D^*}}{ds} + \frac{d \Gamma_{\!X}}{ds}
\end{eqnarray}
 we can then obtain the spectra needed in (\ref{genconv}) using
\begin{eqnarray}\label{final1}
\frac{d \Gamma_{\!D}}{ds} &=& \Gamma_{\!D} \delta(s-m_D^2)\\
\frac{d \Gamma_{\!D^*}}{ds} &=&  \Gamma_{\!D^*} \delta(s-m_{D^*}^2)\\
\frac{d \Gamma_{\!X}}{ds} &=& \frac{d \Gamma}{ds} 
- \frac{d \Gamma_{\!D}}{ds} - \frac{d \Gamma_{\!D^*}}{ds}
\end{eqnarray}
where $\Gamma_{\!D^{(*)}} = \Gamma_{\!D^{(*)}}(\Ecut)$ are the decay
rates for $B \to {D^{(*)}} \ell \bar\nu$ in the presence of a lepton
energy cut. Note that only for $\Ecut=0$ can the values of
$\Gamma_{\!D^{(*)}}$ be obtained from the measured branching
fractions.

As explained above, the theoretical expression for the differential
decay rate is given in terms of a singular expansion. Thus, this
spectrum can only be compared with the experimentally measured
spectrum after smearing it with some appropriate weight function.  One
can show that this weight function has to be a smooth function with
width of order $\sqrt{m_b \lqcd}$. While the detector resolution
function does provide a smearing of the theoretical spectrum, its
width is only several hundred MeV \cite{babarsh}. Thus, additional
smearing is required to compare the calculated spectrum with the
measured spectrum. The simplest observables to compare are moments
of the reconstructed hadron invariant mass spectrum.

These moments can easily be obtained from the general expression given
in Eq.~(\ref{genconv})
\begin{eqnarray}
\langle (s_R - \mDbar^2)^N \rangle \!\!\!&\equiv&\!\!\!
\frac{1}{\Gamma}\int \! ds_R \, (s_R - \mDbar^2)^N 
\frac{d\Gamma}{d s_R} \nn\\ 
\!\!\!&=&\!\!\! \frac{1}{\Gamma}\int \! ds_T \left[
  P^N_D(s_T) \frac{d\Gamma_{\!D}}{d s_T} + P^N_{D^*}(s_T)
\frac{d\Gamma_{\!D^*}}{d s_T}\right.\nn\\ 
&& \left.\qquad\quad + P^N_X(s_T) \frac{d\Gamma_{\!X}}{d s_T} \right]\,.
\end{eqnarray}
Here we have defined 
\begin{eqnarray}
P^N(s_T) = \int \! ds_R \, (s_R-\mDbar^2)^N P(s_R-s_T,s_T)\,,
\end{eqnarray}
where $P$ stands for any of the three resolution functions.  Thus,
moments of the reconstructed spectrum are given by the true spectrum,
weighted with a function which is determined by the detector
response. 

The result in Eq.~(\ref{final1}) can be further simplified by
expanding the functions $P^N(s_T)$ in a Taylor series:
\begin{equation}\label{Pexpand}
P^N(s_T) =\sum_{n}
(s_T-\mDbar^2)^n P^N_{n}
\end{equation}
The coefficients $P^N_n$ depend on the precise form of the detector
resolution function, but we can obtain some estimates using the
expected form of the detector resolution function. We expect that for
a given value of $s_T$ the function $P(\omega,s_T)$ is peaked around
$\omega=0$ with some width $\sigma$ that may depend weakly on
$s_T$. The resolution functions used by the Babar collaboration
\cite{babarsh} have width of order $\sigma \sim 1 \,{\rm GeV}^2$,
while for a measurement using fully reconstructed $B$ events we expect
$\sigma \sim (700 \,{\rm MeV})^2$. We furthermore assume that the
variations in the resolution function are slow compared to the width
of the resolution function. If the width $\sigma$ varies by a small
fraction $\epsilon$ over the range of the hadronic mass,
\begin{equation}
\frac{d^n\sigma}{ds_T^n}\sim \frac{\epsilon \sigma}{(s_T^{\rm max}-s_T^{\rm
min})^n}\sim\epsilon \frac{\sigma}{(m_b^2)^n}
\end{equation}
then the coefficients $P^N_n$ are suppressed by $\epsilon$ for $n>N$:
\begin{equation}\label{estimates}
P^N_n\sim\begin{cases}
\sigma^{N-n}& \text{for $n\le N$}\\
\epsilon \left(\frac\sigma{m_b^2}\right)\frac1{(m_b^2)^{n-N}}&
 \text{for $n>N$}
\end{cases}
\end{equation}

We can now see how the expected form of the resolution function
results in a simple organization of the
expansion~(\ref{Pexpand}). Using the estimates~(\ref{estimates})
and $\langle (s_T-\mDbar^2)^n \rangle \sim
\lqcd^{2n}$ we estimate the different terms in Eq.~(\ref{Pexpand})
as follows:
\begin{multline}\label{RTrelation2}
\langle (s_R - \mDbar^2)^N \rangle \sim \sum_{n=0}^N
\sigma^{N-n}\lqcd^{2n}\\ +\sum_{n>N}\epsilon\frac{\sigma}{m_b^2}
\left(\frac{\lqcd^2}{m_b^2}\right)^{n-N}\lqcd^{2N} 
\end{multline}

Thus, if the resolution functions behave as expected, the terms with
$n\ge N+1$ in the relation~(\ref{RTrelation}) giving the measured
moments in terms of the theory moments are suppressed by
$\epsilon\frac{\sigma}{m_b^2}(\frac{\lqcd^2}{m_b^2})$. Note also that
since present experiments have $\sigma$ substantially larger than
$\lqcd$ the contributions of low $n$ theory moments to the measured
$N$-th moment are amplified by powers of $\sigma/\lqcd^2$. Neglecting the
suppressed terms ($n\ge N+1$) we
thus obtain
\begin{eqnarray}\label{sRsTrelation}
\left\langle (s_R - \mDbar^2)^N \right\rangle &=& \sum_{n=0}^N \left[ 
\frac{\Gamma_D}{\Gamma} 
\left( P_{D,n}^N - P_{X,n}^N \right) \left(m_D^2-\mDbar^2\right)^n
\right.\nn\\
&&\left.
+\frac{\Gamma_{D^*}}{\Gamma} \left( P_{D^*,n}^N - P_{X,n}^N \right) 
\left(m_{D^*}^2-\mDbar^2\right)^n
\right.\nn\\
&&\left.
+ P_{X,n}^N \langle (s_T-\mDbar^2)^n \rangle \right]\,.
\end{eqnarray}
This is the main result of this paper and shows how moments of the
reconstructed hadronic invariant mass are related to the moments of
the true mass distribution. The only ingredient in this relation are
the first few moments of the detector resolution functions. 

As an example of these simplifications we consider a simple model for
the detector resolution function
\begin{eqnarray}
\tilde P(\omega,s_T) = \frac{1}{\sqrt{2\pi} \Delta(s_T)} \exp\left[
-\frac{\omega^2}{2\Delta(s_T)^2}\right]
\end{eqnarray}
Here, the width of the Gaussian distribution depends on the value of
$s_T$, however the variation is slow compared to the width of the
Gaussian itself
\begin{eqnarray}
\frac{d^n}{ds_T^n}\Delta(s_T)\sim\epsilon \frac{\Delta(\mDbar^2)}{m_b^{2n}}
\end{eqnarray}
For this model the Taylor expansions of the first few functions $P^N$,
Eq.~(\ref{Pexpand}), are
\begin{equation}
\begin{aligned}
P^0(s_T) =& 1\\
P^1(s_T) =& (s_T-\mDbar^2)\\
P^2(s_T) =& (s_T-\mDbar^2)^2 + \Delta(\mDbar^2)^2\\
=& \Delta(\mDbar^2)^2\\
+&2\, \Delta(\mDbar^2)\Delta'(\mDbar^2)(s_T-\mDbar^2)\\
+&\left[1+(\Delta'(\mDbar^2))^2
+\Delta(\mDbar^2)\Delta''(\mDbar^2)\right](s_T-\mDbar^2)^2 \\ 
+&\left[\Delta'(\mDbar^2)\Delta''(\mDbar^2)
+\frac{\Delta(\mDbar^2)\Delta'''(\mDbar^2)}{3}\right](s_T-\mDbar^2)^3 \\
&\qquad+{\cal O}\left[(s_T-\mDbar^2)^4\right]
\end{aligned}
\end{equation}
The first two moments are quite trivial, but the second displays
already all of the features of the general organization principles
given in (\ref{estimates}).

To summarize, we give a detailed description of the steps necessary
for a measurement of the first moment of the reconstructed mass
distribution. The procedure for higher moments should be obvious from
this. For each of the detector resolution functions, calculate its
first moment for various values of the true invariant mass $s_T$. This
gives the function $P^1(s_T)$. Fit this function to a third order
polynomial in $(s_T-\mDbar^2)$ and call the coefficients $P^1_n$. The
theoretical expression for the reconstructed moment is then obtained
using Eq.~(\ref{sRsTrelation}) for $N=1$ and using the theoretical
expression for the $\langle (s_T-\mDbar^2)^n \rangle$ given in the Appendix.

For many applications the detector resolution function $P(s_R-s_T,s_T)$
does not depend on the value of the true invariant mass and is
therefore only a function of the difference of the true and the
reconstructed mass
\begin{eqnarray}
P(s_R-s_T,s_T) = p(s_R-s_T)\,.
\end{eqnarray}
In this case, the expression in Eq.~(\ref{sRsTrelation}) simplifies
considerably, since the quantities $P^N_n$ can now be expressed solely
in terms of moments of the resolution function. We find
\begin{eqnarray}
P^N_n =  \binom{N}{n} p^{N-n}\,,
\end{eqnarray}
where $p^N = \int \!ds \, s^N p(s)$. 
This allows to write an even simpler relations between the  moment
of the reconstructed and  true spectra. For example, for the first moment
\begin{eqnarray}\label{prel}
\left\langle s_R - \mDbar^2 \right\rangle &=&
\left\langle s_T - \mDbar^2 \right\rangle
+ p^1_X
\\
&+&
\left( p^1_D - p^1_X \right) \frac{\Gamma_D}{\Gamma} 
+ \left( p^1_{D^*} - p^1_X \right) \frac{\Gamma_{D^*}}{\Gamma}\,,\nn
\end{eqnarray}
Note that Eq.~(\ref{prel}) is exact, i.e., it does not contain unknown
$\epsilon\sigma\lqcd^2/m_b^4$ corrections as Eq.~(\ref{sRsTrelation})
does.  In general, for the $N$-th moment of the reconstructed spectrum
the first $N$ moments of the resolution function, as well as the first
$N$ moments of the true spectrum are needed. For distributions with
vanishing first moment, $p^1=0$, we see that the detector resolution
function introduces no bias into the determination of the first moment
of $ (s_T - \mDbar^2) $. In practice, however, a detector always can
miss particles and the first moments, $p^1$, are therefore expected to
be negative.

To summarize, we have shown how to relate moments of the reconstructed
hadronic invariant mass spectrum to the moments of the true,
calculable, spectrum, involving only moments of the detector
resolution functions. We have given expressions for the general case,
in which the resolution functions can depend on the true invariant
mass, as well as for the special case, in which the resolution
functions only depend on the difference between the true and the
reconstructed invariant mass.

We are indebted to Michael Luke for providing us with code to
calculate perturbative corrections to the hadronic invariant mass
moments. We furthermore would like to thank Oliver Buchm{\"u}ller,
Henning Fl{\"a}cher and Vera Luth for helpful discussion. This work
was supported by the DOE under grant DOE-FG03-97ER40546.

%%%%%%%%%%%%%%%%%%%%%%%%%%%%%%%%%%%%%%%%%%%%%%%%%
%%%%%%%%%%%%%%%%%%%%%%%%%%%%%%%%%%%%%%%%%%%%%%%%%
%%%%%%%%%%%%%%%%%%%%%%%%%%%%%%%%%%%%%%%%%%%%%%%%%
%%%%%%%%%%%%%%%%%%%%%%%%%%%%%%%%%%%%%%%%%%%%%%%%%

\begin{widetext}
\appendix

\section{The first three hadronic moments}

In this appendix we give the expressions for the first three hadronic
moments $S_n = \langle (s_T-\bar m_D^2)^n \rangle$ [note that this
definition of $S_n$ is slightly different than the one for ${\cal
M}_n$ in Eq.~(\ref{Mn})]. Since the hadronic moments satisfy $S_n \sim
(\Lambda_{\rm QCD}/m_B)^n$, the first three moments are non-vanishing
to order $1/m_B^3$.  The decay rate $\Gamma$, and the first two
moments $S_1$, $S_2$ were calculated in the pole scheme as a function
of the lepton energy cut in \cite{fls}. In \cite{bllm} this
calculation was repeated in several different mass schemes and
explicit expressions were given for the dependence on the lepton
energy cut. For completeness, we will repeat these results here and
also give the result for the third moment $S_3$, which has not been
presented in the literature so far. All these results are given in the
1S scheme, and for comparison with earlier works also in the pole scheme. They are written in the form 
\beqa\label{expdef} 
S_n(E_0) &=&
S_n^{(1)}(E_0) 
+ S_n^{(2)}(E_0)\, \Lambda 
+ S_n^{(3)}(E_0)\, \Lambda^2
+S_n^{(4)}(E_0)\, \Lambda^3 
+S_n^{(5)}(E_0)\, \l1 
+S_n^{(6)}(E_0)\, \l2 
\nn\\* && 
+S_n^{(7)}(E_0)\, \l1 \Lambda 
+S_n^{(8)}(E_0)\,
\l2 \Lambda +S_n^{(9)}(E_0)\, \r1 
+S_n^{(10)}(E_0)\, \r2 
+S_n^{(11)}(E_0)\, \t1 
\nn\\* && 
+ S_n^{(12)}(E_0)\, \t2 +S_n^{(13)}(E_0)\, \t3 
+S_n^{(14)}(E_0)\, \t4 
+S_n^{(15)}(E_0)\, \epsilon
+ S_n^{(16)}(E_0)\, \epsilon^2_{\rm BLM}
+ S_n^{(17)}(E_0)\, \epsilon \Lambda \,, 
\eeqa 
where $E_0$ is the value of the lepton
energy cut, and the coefficients $S_n^{(m)}(E_0)$ has the appropriate mass dimension. The parameter $\epsilon\equiv 1$ and $\epsilon_{\rm BLM}\equiv 1$
denote the order in perturbation theory, for which we have used $\alpha_s(m_b) = 0.22$ and $\beta_0 = 25/3$. The definition of the parameter $\Lambda$ depends on the mass scheme used and is
\begin{eqnarray}
\Lambda^{\rm pole} &=& \bar m_B - m_b^{\rm pole} + \frac{\lambda_1}{2m_b} - \frac{\rho_1 - {\cal T}_1 - {\cal T}_3}{m_b^2}\\
\Lambda^{\rm 1S} &=& \frac{m_\Upsilon}{2} - m_b^{\rm 1S}\,.
\end{eqnarray}
The values for the
coefficients $S_n^{(i)}$ are shown in tables I to VI. 
The tables do not contain the $\epsilon \Lambda$ term for non-zero
lepton energy cut $E_0$. This is because the radiative corrections to
the hadronic energy distribution with a lepton energy cut are not
known; for details see Ref.~\cite{fl}. 
In Appendix B we give analytical  expressions required for the
$\epsilon \Lambda$ term with zero lepton energy cut.

\begingroup\squeezetable
\begin{table*}[t]
\caption{Coefficients for $S_1(E_0)$ in the Pole scheme as a function of $E_0$.}
\begin{tabular}{c||ccccccccccccccccc}
$E_0$  &  $S_1^{(1)}$  &  $S_1^{(2)}$  &  $S_1^{(3)}$  &  $S_1^{(4)}$  &  
  $S_1^{(5)}$  &  $S_1^{(6)}$  &  $S_1^{(7)}$  &  $S_1^{(8)}$  &  
  $S_1^{(9)}$  &  $S_1^{(10)}$  &  $S_1^{(11)}$  &  $S_1^{(12)}$  &  
  $S_1^{(13)}$  &  $S_1^{(14)}$  &  $S_1^{(15)}$  &  $S_1^{(16)}$  &  
  $S_1^{(17)}$  \\ \hline\hline
$0$ & $0
$ & $1.248
$ & $0.262
$ & $0.06
$ & $1.02
$ & $-0.32
$ & $0.41
$ & $-0.11
$ & $0.42
$ & $-0.21
$ & $0.3
$ & $0.15
$ & $0.28
$ & $0.08
$ & $0.102
$ & $0.111
$ & $0.038
$ \\

$0.5$ & $0
$ & $1.231
$ & $0.258
$ & $0.06
$ & $1.03
$ & $-0.29
$ & $0.41
$ & $-0.09
$ & $0.43
$ & $-0.21
$ & $0.31
$ & $0.16
$ & $0.28
$ & $0.08
$ & $0.097
$ & $0.107
$ & - \\

$0.7$ & $0
$ & $1.209
$ & $0.251
$ & $0.06
$ & $1.05
$ & $-0.25
$ & $0.42
$ & $-0.07
$ & $0.45
$ & $-0.21
$ & $0.31
$ & $0.17
$ & $0.28
$ & $0.09
$ & $0.092
$ & $0.102
$ & - \\

$0.9$ & $0
$ & $1.18
$ & $0.241
$ & $0.06
$ & $1.08
$ & $-0.18
$ & $0.44
$ & $-0.03
$ & $0.48
$ & $-0.2
$ & $0.31
$ & $0.18
$ & $0.29
$ & $0.1
$ & $0.084
$ & $0.095
$ & - \\

$1.1$ & $0
$ & $1.148
$ & $0.228
$ & $0.05
$ & $1.13
$ & $-0.08
$ & $0.46
$ & $0.03
$ & $0.54
$ & $-0.19
$ & $0.32
$ & $0.21
$ & $0.29
$ & $0.12
$ & $0.075
$ & $0.086
$ & - \\

$1.3$ & $0
$ & $1.118
$ & $0.211
$ & $0.04
$ & $1.22
$ & $0.04
$ & $0.51
$ & $0.12
$ & $0.63
$ & $-0.16
$ & $0.34
$ & $0.26
$ & $0.3
$ & $0.14
$ & $0.064
$ & $0.077
$ & - \\

$1.5$ & $0
$ & $1.1
$ & $0.194
$ & $0.04
$ & $1.38
$ & $0.2
$ & $0.6
$ & $0.26
$ & $0.8
$ & $-0.11
$ & $0.37
$ & $0.35
$ & $0.31
$ & $0.17
$ & $0.054
$ & $0.067
$ & - \\

$1.7$ & $0
$ & $1.112
$ & $0.188
$ & $0.03
$ & $1.7
$ & $0.4
$ & $0.83
$ & $0.47
$ & $1.16
$ & $0.04
$ & $0.43
$ & $0.53
$ & $0.32
$ & $0.21
$ & $0.044
$ & $0.057
$ & -

\end{tabular}
\end{table*}
\endgroup

\begingroup\squeezetable
\begin{table*}[t]
\caption{Coefficients for $S_2(E_0)$ in the Pole scheme as a function of $E_0$.}
\begin{tabular}{c||ccccccccccccccccc}
$E_0$  &  $S_2^{(1)}$  &  $S_2^{(2)}$  &  $S_2^{(3)}$  &  $S_2^{(4)}$  &  
  $S_2^{(5)}$  &  $S_2^{(6)}$  &  $S_2^{(7)}$  &  $S_2^{(8)}$  &  
  $S_2^{(9)}$  &  $S_2^{(10)}$  &  $S_2^{(11)}$  &  $S_2^{(12)}$  &  
  $S_2^{(13)}$  &  $S_2^{(14)}$  &  $S_2^{(15)}$  &  $S_2^{(16)}$  &  
  $S_2^{(17)}$  \\ \hline\hline
$0$ & $0
$ & $0
$ & $1.853
$ & $0.76
$ & $-3.9
$ & $0
$ & $1.69
$ & $-1.63
$ & $-4.52
$ & $1.24
$ & $-0.73
$ & $-2.2
$ & $0
$ & $0
$ & $0.301
$ & $0.255
$ & $0.4
$ \\

$0.5$ & $0
$ & $0
$ & $1.808
$ & $0.74
$ & $-3.84
$ & $0
$ & $1.72
$ & $-1.53
$ & $-4.57
$ & $1.16
$ & $-0.72
$ & $-2.17
$ & $0
$ & $0
$ & $0.273
$ & $0.235
$ & - \\

$0.7$ & $0
$ & $0
$ & $1.756
$ & $0.71
$ & $-3.77
$ & $0
$ & $1.76
$ & $-1.39
$ & $-4.66
$ & $1.04
$ & $-0.71
$ & $-2.13
$ & $0
$ & $0
$ & $0.241
$ & $0.212
$ & - \\

$0.9$ & $0
$ & $0
$ & $1.69
$ & $0.67
$ & $-3.67
$ & $0
$ & $1.84
$ & $-1.19
$ & $-4.83
$ & $0.87
$ & $-0.69
$ & $-2.07
$ & $0
$ & $0
$ & $0.202
$ & $0.182
$ & - \\

$1.1$ & $0
$ & $0
$ & $1.62
$ & $0.63
$ & $-3.56
$ & $0
$ & $1.99
$ & $-0.94
$ & $-5.1
$ & $0.65
$ & $-0.67
$ & $-2.01
$ & $0
$ & $0
$ & $0.16
$ & $0.149
$ & - \\

$1.3$ & $0
$ & $0
$ & $1.557
$ & $0.58
$ & $-3.46
$ & $0
$ & $2.25
$ & $-0.63
$ & $-5.56
$ & $0.37
$ & $-0.65
$ & $-1.95
$ & $0
$ & $0
$ & $0.12
$ & $0.115
$ & - \\

$1.5$ & $0
$ & $0
$ & $1.516
$ & $0.53
$ & $-3.4
$ & $0
$ & $2.69
$ & $-0.24
$ & $-6.39
$ & $0.02
$ & $-0.64
$ & $-1.92
$ & $0
$ & $0
$ & $0.083
$ & $0.084
$ & - \\

$1.7$ & $0
$ & $0
$ & $1.525
$ & $0.52
$ & $-3.43
$ & $0
$ & $3.51
$ & $0.27
$ & $-8.05
$ & $-0.43
$ & $-0.65
$ & $-1.94
$ & $0
$ & $0
$ & $0.051
$ & $0.056
$ & -

\end{tabular}
\end{table*}
\endgroup

\begingroup\squeezetable
\begin{table*}[t]
\caption{Coefficients for $S_3(E_0)$ in the Pole scheme as a function of $E_0$.}
\begin{tabular}{c||ccccccccccccccccc}
$E_0$  &  $S_3^{(1)}$  &  $S_3^{(2)}$  &  $S_3^{(3)}$  &  $S_3^{(4)}$  &  
  $S_3^{(5)}$  &  $S_3^{(6)}$  &  $S_3^{(7)}$  &  $S_3^{(8)}$  &  
  $S_3^{(9)}$  &  $S_3^{(10)}$  &  $S_3^{(11)}$  &  $S_3^{(12)}$  &  
  $S_3^{(13)}$  &  $S_3^{(14)}$  &  $S_3^{(15)}$  &  $S_3^{(16)}$  &  
  $S_3^{(17)}$  \\ \hline\hline
$0$ & $0
$ & $0
$ & $0
$ & $3.0
$ & $0
$ & $0
$ & $-17.63
$ & $0
$ & $21.26
$ & $0
$ & $0
$ & $0
$ & $0
$ & $0
$ & $1.669
$ & $1.257
$ & $1.693
$ \\

$0.5$ & $0
$ & $0
$ & $0
$ & $2.9
$ & $0
$ & $0
$ & $-17.18
$ & $0
$ & $20.88
$ & $0
$ & $0
$ & $0
$ & $0
$ & $0
$ & $1.423
$ & $1.097
$ & - \\

$0.7$ & $0
$ & $0
$ & $0
$ & $2.8
$ & $0
$ & $0
$ & $-16.65
$ & $0
$ & $20.41
$ & $0
$ & $0
$ & $0
$ & $0
$ & $0
$ & $1.171
$ & $0.924
$ & - \\

$0.9$ & $0
$ & $0
$ & $0
$ & $2.67
$ & $0
$ & $0
$ & $-16.
$ & $0
$ & $19.81
$ & $0
$ & $0
$ & $0
$ & $0
$ & $0
$ & $0.888
$ & $0.72
$ & - \\

$1.1$ & $0
$ & $0
$ & $0
$ & $2.53
$ & $0
$ & $0
$ & $-15.32
$ & $0
$ & $19.15
$ & $0
$ & $0
$ & $0
$ & $0
$ & $0
$ & $0.619
$ & $0.52
$ & - \\

$1.3$ & $0
$ & $0
$ & $0
$ & $2.42
$ & $0
$ & $0
$ & $-14.7
$ & $0
$ & $18.56
$ & $0
$ & $0
$ & $0
$ & $0
$ & $0
$ & $0.394
$ & $0.345
$ & - \\

$1.5$ & $0
$ & $0
$ & $0
$ & $2.33
$ & $0
$ & $0
$ & $-14.3
$ & $0
$ & $18.18
$ & $0
$ & $0
$ & $0
$ & $0
$ & $0
$ & $0.222
$ & $0.205
$ & - \\

$1.7$ & $0
$ & $0
$ & $0
$ & $2.33
$ & $0
$ & $0
$ & $-14.36
$ & $0
$ & $18.34
$ & $0
$ & $0
$ & $0
$ & $0
$ & $0
$ & $0.105
$ & $0.104
$ & -

\end{tabular}
\end{table*}
\endgroup

\begingroup\squeezetable
\begin{table*}[t]
\caption{Coefficients for $S_1(E_0)$ in the $1S$ scheme as a function of $E_0$.}
\begin{tabular}{c||ccccccccccccccccc}
$E_0$  &  $S_1^{(1)}$  &  $S_1^{(2)}$  &  $S_1^{(3)}$  &  $S_1^{(4)}$  &  
  $S_1^{(5)}$  &  $S_1^{(6)}$  &  $S_1^{(7)}$  &  $S_1^{(8)}$  &  
  $S_1^{(9)}$  &  $S_1^{(10)}$  &  $S_1^{(11)}$  &  $S_1^{(12)}$  &  
  $S_1^{(13)}$  &  $S_1^{(14)}$  &  $S_1^{(15)}$  &  $S_1^{(16)}$  &  
  $S_1^{(17)}$  \\ \hline\hline
$0$ & $0.832
$ & $1.633
$ & $0.416
$ & $0.13
$ & $1.49
$ & $-0.36
$ & $0.75
$ & $0.
$ & $0.46
$ & $-0.24
$ & $0.53
$ & $0.25
$ & $0.5
$ & $0.14
$ & $0.044
$ & $-0.025
$ & $0.025$
\\

$0.5$ & $0.82
$ & $1.609
$ & $0.409
$ & $0.12
$ & $1.5
$ & $-0.32
$ & $0.75
$ & $0.02
$ & $0.48
$ & $-0.24
$ & $0.54
$ & $0.26
$ & $0.5
$ & $0.14
$ & $0.039
$ & $-0.028
$ & -
\\

$0.7$ & $0.805
$ & $1.578
$ & $0.398
$ & $0.12
$ & $1.52
$ & $-0.26
$ & $0.77
$ & $0.05
$ & $0.5
$ & $-0.23
$ & $0.54
$ & $0.27
$ & $0.5
$ & $0.16
$ & $0.032
$ & $-0.031
$ & -
\\

$0.9$ & $0.784
$ & $1.533
$ & $0.38
$ & $0.11
$ & $1.56
$ & $-0.16
$ & $0.79
$ & $0.12
$ & $0.55
$ & $-0.22
$ & $0.55
$ & $0.3
$ & $0.51
$ & $0.18
$ & $0.023
$ & $-0.035
$ & -
\\

$1.1$ & $0.759
$ & $1.479
$ & $0.354
$ & $0.1
$ & $1.63
$ & $-0.02
$ & $0.83
$ & $0.22
$ & $0.63
$ & $-0.2
$ & $0.57
$ & $0.34
$ & $0.52
$ & $0.2
$ & $0.011
$ & $-0.04
$ & -
\\

$1.3$ & $0.734
$ & $1.42
$ & $0.319
$ & $0.09
$ & $1.74
$ & $0.18
$ & $0.91
$ & $0.38
$ & $0.77
$ & $-0.16
$ & $0.59
$ & $0.41
$ & $0.54
$ & $0.24
$ & $-0.002
$ & $-0.046
$ & -
\\

$1.5$ & $0.716
$ & $1.371
$ & $0.277
$ & $0.06
$ & $1.97
$ & $0.45
$ & $1.07
$ & $0.65
$ & $1.03
$ & $-0.06
$ & $0.64
$ & $0.55
$ & $0.56
$ & $0.3
$ & $-0.018
$ & $-0.054
$ & -
\\

$1.7$ & $0.72
$ & $1.368
$ & $0.254
$ & $0.05
$ & $2.49
$ & $0.84
$ & $1.59
$ & $1.13
$ & $1.64
$ & $0.22
$ & $0.76
$ & $0.86
$ & $0.6
$ & $0.38
$ & $-0.035
$ & $-0.066
$ & -
\\

\end{tabular}
\end{table*}
\endgroup

\begingroup\squeezetable
\begin{table*}[t]
\caption{Coefficients for $S_2(E_0)$ in the $1S$ scheme as a function of $E_0$.}
\begin{tabular}{c||ccccccccccccccccc}
$E_0$  &  $S_2^{(1)}$  &  $S_2^{(2)}$  &  $S_2^{(3)}$  &  $S_2^{(4)}$  &  
  $S_2^{(5)}$  &  $S_2^{(6)}$  &  $S_2^{(7)}$  &  $S_2^{(8)}$  &  
  $S_2^{(9)}$  &  $S_2^{(10)}$  &  $S_2^{(11)}$  &  $S_2^{(12)}$  &  
  $S_2^{(13)}$  &  $S_2^{(14)}$  &  $S_2^{(15)}$  &  $S_2^{(16)}$  &  
  $S_2^{(17)}$  \\ \hline\hline
$0$ & $0.816
$ & $3.188
$ & $3.889
$ & $1.73
$ & $-1.95
$ & $-1.28
$ & $5.06
$ & $-2.78
$ & $-4.69
$ & $0.67
$ & $-0.06
$ & $-2.38
$ & $0.77
$ & $0.1
$ & $0.454
$ & $-0.47
$ & $0.356
$ 
\\

$0.5$ & $0.795
$ & $3.106
$ & $3.784
$ & $1.68
$ & $-1.88
$ & $-1.19
$ & $5.08
$ & $-2.55
$ & $-4.75
$ & $0.57
$ & $-0.05
$ & $-2.32
$ & $0.76
$ & $0.11
$ & $0.469
$ & $-0.465
$ & -
\\

$0.7$ & $0.77
$ & $3.005
$ & $3.651
$ & $1.61
$ & $-1.79
$ & $-1.06
$ & $5.13
$ & $-2.23
$ & $-4.83
$ & $0.43
$ & $-0.03
$ & $-2.23
$ & $0.76
$ & $0.13
$ & $0.499
$ & $-0.458
$ & -
\\

$0.9$ & $0.738
$ & $2.872
$ & $3.471
$ & $1.5
$ & $-1.63
$ & $-0.87
$ & $5.24
$ & $-1.74
$ & $-4.99
$ & $0.24
$ & $-0.01
$ & $-2.11
$ & $0.75
$ & $0.16
$ & $0.562
$ & $-0.446
$ & -
\\

$1.1$ & $0.702
$ & $2.722
$ & $3.258
$ & $1.37
$ & $-1.42
$ & $-0.62
$ & $5.48
$ & $-1.08
$ & $-5.24
$ & $-0.02
$ & $0.03
$ & $-1.94
$ & $0.74
$ & $0.2
$ & $0.678
$ & $-0.43
$ & -
\\

$1.3$ & $0.668
$ & $2.572
$ & $3.031
$ & $1.2
$ & $-1.12
$ & $-0.3
$ & $5.94
$ & $-0.21
$ & $-5.66
$ & $-0.34
$ & $0.08
$ & $-1.72
$ & $0.74
$ & $0.26
$ & $0.888
$ & $-0.411
$ & -
\\

$1.5$ & $0.641
$ & $2.45
$ & $2.826
$ & $1.03
$ & $-0.68
$ & $0.11
$ & $6.9
$ & $0.98
$ & $-6.41
$ & $-0.7
$ & $0.17
$ & $-1.43
$ & $0.76
$ & $0.34
$ & $1.302
$ & $-0.388
$ & -
\\

$1.7$ & $0.638
$ & $2.423
$ & $2.746
$ & $0.92
$ & $0.13
$ & $0.72
$ & $9.27
$ & $2.84
$ & $-7.99
$ & $-1.03
$ & $0.35
$ & $-0.94
$ & $0.82
$ & $0.47
$ & $2.295
$ & $-0.369
$ & -
\\

\end{tabular}
\end{table*}
\endgroup

\begingroup\squeezetable
\begin{table*}[t]
\caption{Coefficients for $S_3(E_0)$ in the $1S$ scheme as a function of $E_0$.}
\begin{tabular}{c||ccccccccccccccccc}
$E_0$  &  $S_3^{(1)}$  &  $S_3^{(2)}$  &  $S_3^{(3)}$  &  $S_3^{(4)}$  &  
  $S_3^{(5)}$  &  $S_3^{(6)}$  &  $S_3^{(7)}$  &  $S_3^{(8)}$  &  
  $S_3^{(9)}$  &  $S_3^{(10)}$  &  $S_3^{(11)}$  &  $S_3^{(12)}$  &  
  $S_3^{(13)}$  &  $S_3^{(14)}$  &  $S_3^{(15)}$  &  $S_3^{(16)}$  &  
  $S_3^{(17)}$  \\ \hline\hline
$0$ & $0.87
$ & $5.084
$ & $11.102
$ & $11.43
$ & $-9.59
$ & $-2.52
$ & $-12.73
$ & $-10.45
$ & $10.06
$ & $3.84
$ & $-1.51
$ & $-7.75
$ & $1.08
$ & $0.03
$ & $3.094
$ & $0.309
$ & $3.441
$ 
\\

$0.5$ & $0.84
$ & $4.905
$ & $10.7
$ & $10.99
$ & $-9.26
$ & $-2.34
$ & $-12.05
$ & $-9.68
$ & $9.54
$ & $3.47
$ & $-1.45
$ & $-7.5
$ & $1.06
$ & $0.05
$ & $3.067
$ & $0.255
$ & -
\\

$0.7$ & $0.806
$ & $4.7
$ & $10.232
$ & $10.47
$ & $-8.85
$ & $-2.12
$ & $-11.14
$ & $-8.7
$ & $8.82
$ & $3.
$ & $-1.38
$ & $-7.18
$ & $1.04
$ & $0.08
$ & $3.1
$ & $0.205
$ & -
\\

$0.9$ & $0.764
$ & $4.445
$ & $9.638
$ & $9.78
$ & $-8.29
$ & $-1.81
$ & $-9.8
$ & $-7.31
$ & $7.75
$ & $2.35
$ & $-1.28
$ & $-6.76
$ & $1.01
$ & $0.12
$ & $3.235
$ & $0.148
$ & -
\\

$1.1$ & $0.719
$ & $4.167
$ & $8.977
$ & $8.99
$ & $-7.61
$ & $-1.43
$ & $-8.02
$ & $-5.55
$ & $6.3
$ & $1.56
$ & $-1.16
$ & $-6.26
$ & $0.98
$ & $0.17
$ & $3.558
$ & $0.1
$ & -
\\

$1.3$ & $0.675
$ & $3.896
$ & $8.314
$ & $8.15
$ & $-6.83
$ & $-0.96
$ & $-5.68
$ & $-3.38
$ & $4.33
$ & $0.62
$ & $-1.02
$ & $-5.7
$ & $0.97
$ & $0.25
$ & $4.223
$ & $0.062
$ & -
\\

$1.5$ & $0.641
$ & $3.674
$ & $7.739
$ & $7.37
$ & $-5.94
$ & $-0.38
$ & $-2.43
$ & $-0.6
$ & $1.52
$ & $-0.47
$ & $-0.84
$ & $-5.09
$ & $0.97
$ & $0.35
$ & $5.608
$ & $0.028
$ & -
\\

$1.7$ & $0.631
$ & $3.594
$ & $7.486
$ & $6.94
$ & $-4.75
$ & $0.44
$ & $3.2
$ & $3.45
$ & $-3.39
$ & $-1.76
$ & $-0.58
$ & $-4.35
$ & $1.04
$ & $0.52
$ & $9.047
$ & $-0.01
$ & -
\\

\end{tabular}
\end{table*}
\endgroup

\section{Analytic expressions for perturbative corrections to new partonic moments}
The leading perturbative corrections to the hadronic invariant moments arise only from bremsstrahlung contributions, since the virtual contributions are proportional to a $\delta$-function which does not contribute to the moments. These bremsstrahlung contributions were calculated in \cite{fl} with a cut on the lepton energy. However, virtual diagrams do contribute to perturbative corrections to the terms supressed by powers of $\lqcd/m_b$. Using
\begin{eqnarray}
s_H = m_B^2 + 2 m_B m_b (\hat e_0-1) + m_b^2(1+\hat s_0-2\hat e_0)\,.
\end{eqnarray}
the corrections proportional to $\alpha_s \lqcd/m_b$ can be obtained from the perturbative corrections to  the partonic invariant mass $\hat s_0 = p_c^2/m_b^2$ and partonic energy
$\hat e_0 = e_c/m_b$. For the first two hadronic invariant mass moments one needs the partonic
moments $M^{(0,0)}$, $M^{(1,0)}$, $M^{(2,0)}$, $M^{(0,1)}$,
$M^{(0,2)}$ and $M^{(1,1)}$, where we used the definition of
\cite{fls}
\begin{eqnarray}
M^{(n,m)} = \int \!\! d\hat s_0 \,d\hat e_0 \,\, 
(\hat s_0-\mqhat^2)^n e_0^m \frac{d\Gamma}{d\hat s_0 d\hat e_0},
\end{eqnarray}
with $\mqhat=m_c/m_b$. 
Perturbative corrections to these partonic moments can be obtained using the results of \cite{perturbative} and analytical results were given in \cite{fls}. Here we give the 
expressions for the additional moments needed to calculate the third
hadronic invariant mass moments. Writing
\begin{eqnarray}
M^{(n,m)}  = M_{\rm non-pert}^{(n,m)} + \frac{\alpha_s}{\pi} M_{\rm pert}^{(n,m)}
\end{eqnarray}
we find for the perturbative corrections
\begin{eqnarray}
M_{\rm pert}^{(3,0)} &=& \frac{377}{132300} - \frac{119}{2700} \,\mqhat^2 
+ \frac{401}{900} \,\mqhat^4 + \frac{97}{36} \,\mqhat^6 
- \frac{1}{12} \,\mqhat^8 - \frac{491}{900} \,\mqhat^{10} 
- \frac{5531}{2700} \,\mqhat^{12} - \frac{55747}{132300} \,\mqhat^{14} \nn\\
&&+ \left( \frac{16}{9} \, \mqhat^6 + \frac{29}{9} \, \mqhat^8 
+ \frac{43}{15} \, \mqhat^{10} + \frac{7}{5} \, \mqhat^{12} 
+ \frac{11}{105} \, \mqhat^{14}  \right) \log(\mqhat^2)\\
M_{\rm pert}^{(2,1)} &=&\frac{1081}{132300} - \frac{187}{1350} \,\mqhat^2 
- \frac{463}{600} \,\mqhat^4 + \frac{127}{216} \,\mqhat^6 
- \frac{1}{12} \,\mqhat^8 + \frac{67}{150} \,\mqhat^{10} 
- \frac{259}{1800} \,\mqhat^{12} + \frac{25033}{264600} \,\mqhat^{14} \nn\\
&&- \left( \frac{7}{12} \, \mqhat^4 + \frac{7}{12} \, \mqhat^6 
+ \frac{4}{9} \, \mqhat^8 + \frac{1}{15} \, \mqhat^{10} 
+ \frac{13}{180} \, \mqhat^{12} + \frac{13}{420} \, \mqhat^{14}  \right) 
\log(\mqhat^2) \\
M_{\rm pert}^{(1,2)} &=& \frac{7421}{176400} + \frac{4079}{10800} \,\mqhat^2 
- \frac{341}{1200} \,\mqhat^4 - \frac{71}{144} \,\mqhat^6 
+ \frac{17}{48} \,\mqhat^8 + \frac{257}{1200} \,\mqhat^{10} 
- \frac{131}{1200} \,\mqhat^{12} - \frac{53819}{529200} \,\mqhat^{14} \nn\\
&&+ \left( \frac{4}{15} \,\mqhat^2  + \frac{13}{60} \, \mqhat^4 
+ \frac{1}{12} \, \mqhat^6 - \frac{17}{36} \, \mqhat^8 
+ \frac{17}{60} \, \mqhat^{10} + \frac{2}{15} \, \mqhat^{12} 
+ \frac{4}{105} \, \mqhat^{14}  \right) \log(\mqhat^2) \\
M_{\rm pert}^{(0,3)} &=& \frac{289223}{1058400} + \frac{762659}{1058400} \, \mqhat^2 
-\frac{67337}{19600} \, \mqhat^4 + \frac{28364999}{529200} \, \mqhat^6 
-  \frac{57005623}{1058400} \, \mqhat^8 + \frac{139623}{39200} \, \mqhat^{10}
 - \frac{21613}{58800} \, \mqhat^{12} 
 \nn\\
&&
- \frac{260423}{529200} \, \mqhat^{14} - \pi^2 \left( \frac{1}{28} +
\frac{7}{60} \, \mqhat^2 + \frac{1}{12} \, \mqhat^4 - \frac{127}{36}
\, \mqhat^6 + \frac{2176}{105} \, \mqhat^7 - \frac{127}{36} \,
 \mqhat^8 + \frac{1}{12} \, \mqhat^{10} + \frac{7}{60} \, \mqhat^{12}
+ \frac{1}{28} \, \mqhat^{14} \right) \nn\\ 
&& + \log(\mqhat^2)
\left( \frac{1}{7} \, \mqhat^2 - \frac{499}{560} \, \mqhat^4 +
 \frac{126121}{5040} \, \mqhat^6 - \frac{8704}{105} \log(1+\mqhat) \,
 \mqhat^7 + \frac{57373}{2520} \, \mqhat^8 - \frac{667}{2520} \,
 \mqhat^{10} \right.\nn\\ 
&&\left.  
- \frac{257}{25200} \, \mqhat^{12}
- \frac{2743}{35280} \, \mqhat^{14}\right) 
- \log(1-\mqhat^2) \left(
 \frac{151}{882} + \frac{41}{225} \, \mqhat^2 - \frac{5}{9} \,
 \mqhat^4 + \frac{3}{2} \, \mqhat^6 - \frac{3}{2} \, \mqhat^8 +
 \frac{5}{9} \, \mqhat^{10} - \frac{41}{225} \, \mqhat^{12}
- \right.\nn\\ 
&&\left.  \frac{151}{882} \, \mqhat^{14}\right) + {\rm
 Li}_2(\mqhat) \frac{17408}{105} \, \mqhat^7 + {\rm Li}_2(\mqhat^2)
 \left( \frac{3}{14} + \frac{7}{10} \, \mqhat^2 + \frac{1}{2} \,
 \mqhat^4 - \frac{127}{6} \, \mqhat^6 - \frac{4352}{105} \, \mqhat^7
- \frac{127}{6} \, \mqhat^8 + \frac{1}{2} \, \mqhat^{10}
- \right.\nn\\ 
&&\left.  + \frac{7}{10} \, \mqhat^{12} + \frac{3}{14}
\, \mqhat^{14}\right) + \log(1-\mqhat^2) \log(\mqhat^2) \left(
 \frac{1}{7} + \frac{7}{15} \, \mqhat^2 + \frac{1}{3} \, \mqhat^4 -
 \frac{65}{3} \, \mqhat^6 + \frac{4352}{105} \, \mqhat^7 -
 \frac{65}{3} \, \mqhat^8 + \frac{1}{3} \, \mqhat^{10} \right.\nn\\
- &&\left.  + \frac{7}{15} \, \mqhat^{12} + \frac{1}{7} \,
 \mqhat^{14}\right) + \log^2(\mqhat^2) \left( \frac{25}{6} \,
 \mqhat^6 + \frac{83}{12} \, \mqhat^8 - \frac{1}{12} \, \mqhat^{10}
- \frac{7}{60} \, \mqhat^{12} - \frac{1}{28} \, \mqhat^{14}\right)
\end{eqnarray}
\end{widetext}

\end{document}